\documentstyle[12pt]{article}
\def \a{\alpha}
\def \b{\beta}
\def \l{\lambda}
\def \d{\delta}
\def \p{\partial}
\def \dg{\dagger}
\def \e{\epsilon}
\def \s{\sigma}
\def \t{\theta}
\def \T{\Theta}

\begin{document}

\centerline{\bf Hamiltonian Analysis of Gauged $CP^1$ Model,}
\centerline{\bf with or without Hopf term, and fractional spin}

\vskip 0.5in

\centerline{B. Chakraborty\footnote{e-mail: biswajit@bose.ernet.in}}

\vskip 0.2in

\centerline{S. N. Bose National Centre for Basic Sciences}

\centerline{Block-JD, Sector-III, Salt Lake, Calcutta-700091, India.}

\vskip 0.5in

Abstract: Recently it has been shown by Cho and Kimm that the gauged $CP^1$
model, obtained by gauging the  global $SU(2)$ group of $CP^1$ model and adding
a corresponding Chern-Simons term, has got its own soliton. These solitons
are somewhat distinct from those of pure $CP^1$ model, as they cannot
always be characterised by $\pi_2(CP^1)=Z$. In this paper, we first carry out
the Hamiltonian analysis of this gauged $CP^1$ model. Then we couple the Hopf
term, associated to these solitons and again carry out its Hamiltonian analysis.
The symplectic structures, along with the structures of the constraints, 
of these two models (with or without Hopf term) are found to be essentially
the same. The model with Hopf term,is then shown to have fractional spin,
which however depends
not only on the soliton number $N$ but also on the nonabelian charge.

\pagebreak

1.{\bf Introduction}

Recently the
re has been an upsurge of interest in the study of physics
of $2+1$-dimensional systems. Particularly because of the strange nature
of the Poincare group $ISO(2,1)$ in $2+1$-dimension, in contrast to
$ISO(3,1)$, there arises possibilities of nontrivial configuration space
${\cal Q}$ and the associated fractional spin and statistics. These
possibilities can be realised in practice by adding topological terms like
the Chern-Simons(CS) or Hopf term in the model[1,2]. Fractional spin
and Galilean/Poincare' symmetry in these various
models have been exhibited in detail in the literature, where both path
integral[2,3,4] and the canonical analysis[2,5-9] have been performed.

The CS term (abelian) is a local expression ($\sim \e^{\mu \nu \l}a_{\mu}\p_{\nu}a_{\l}$)
involving a ``photon-less" gauge field $a_{\mu}$[5].
This gauge field is basically introduced to mimic, in the manner of 
Aharanov-Bohm, the phase acquired by the system in traversing a 
nontrivial loop in the configuration space[2]. On the other hand, the Hopf term
is usually constructed by writing the conserved current $j^{\mu}$
$(\p_{\mu}j^{\mu}=0)$ of a model as a curl of a `fictitious' gauge
field $a_{\mu}$:
$$j^{\mu}=\e^{\mu \nu \l}\p_{\nu}a_{\l}\eqno(1.1)$$
and then contracting $j^{\mu}$ with $a_{\mu}$ to get the Hopf term $H$ as,
$$H \sim \int d^3x j^{\mu}a_{\mu}\eqno(1.2)$$
Written entirely in terms of $a_{\mu}$ (using (1.1)), this Hopf term has
also the appearance of CS term. However there is a subtle difference.
In the case of CS term, the gauge field $a_{\mu}$ should be counted as an
independent variable in the configuration space[7-9]. This is despite the fact
that the gauge field is ``photon-less''. On the other hand, the $a_{\mu}$
in (1.1) is really a `fictitious' gauge field (as we have mentioned above) 
and is not an independent variable in the configuration space. It
has to be, rather, determined by inverting (1.1), by making use of a suitable
gauge fixing condition. Once done that, the Hopf term (1.2) represents 
a non-local current-current interaction. It should however be mentioned
that this distinction, in terminology, has not always been maintained in the
literature (see for example [4]).

This inherent non-locality arising in the very construction of the Hopf term,
can however be avoided in certain cases by enlarging the phase space.
For example, consider the $O(3)$ non-linear sigma model(NLSM). The model has
solitons[10] and an associated topological current. The Hopf term here is
again non-local. On the other hand, one can
consider $CP^1$ model, which is known to be equivalent to NLSM[10,11]. This is
a $U(1)$ gauge theory, having an enlarged phase space. Here the Hopf term
becomes local[11,12]. In this case, the Hopf term has a geometrical interpretation and
provides a representation of $\pi_1({\cal Q})$. Note that the configuration
space for NLSM is basically given as the space ${\cal Q}=Map (S^2,S^2)$, so
that $\pi_1({\cal Q})=\pi_3(S^2)=Z$. However it should be mentioned that
this trick of enlarging the phase space and writing a local
expression of the Hopf term may not work all the time, as we shall see later
in this paper.

That the Hopf term can impart fractional spin, was demonstrated initially
by Wilczek and Zee[3], in the context of the NLSM, using path integral
technique. It has been found to depend on the soliton number. This result
was later corroborated by Bowick et al.[13], using canonical quantization.
On the other hand, the fractional spin obtained in the models involving
abelian(nonabelian) CS term, have been found to depend on the total abelian 
(nonabelian) charge of the system.

This is an important observation, considering the fact that NLSM has become
almost ubiquitious in physics, appearing in various circumstances where the
original $O(3)$ symmetry is broken spontaneously. For example, in particle
physics, the model is considered a prototype of QCD, as the model is
asymptotically free in $(1+1)$ dimension. On the other hand, in condensed
matter physics, this model can describe antiferromagnetic spin chain in its
relativistic version[14]. And in its nonrelativistic version, it can describe a
Heisenberg ferromagnetic system in the long wavelength limit, i.e. the
Landau-Lifshitz(LL) model[15,16]. Besides, the Hopf term can arise naturally in this
NLSM, when one quantises a $U(1)$ degree of freedom hidden in the configuration
space ${\cal Q}$, as has been shown recently by Kobayashi et. al.[17]. Further,
it has been shown recently in [12], that the Hopf term can alter the spin
algebra of the LL model drastically.

This NLSM has global $O(3)$ symmetry. Recently Nardelli[18] has shown that
if this $O(3)$ group is gauged by adding an $SO(3)$ CS term, then the resulting
model also has got its own soliton. This work was later extended by Cho and
Kimm[19] for the general $CP^N$ model, where one has to gauge the global
$SU(N+1)$ group and add a corresponding CS term.
These solitons are somewhat distinct from those of pure
$CP^N$ model, in the sense that these are not always characterised by the second
homotopy group ($\pi_2(CP^N)=Z$) of the manifold, unlike the pure $CP^1$
model[10]. 

The purpose of the paper is to investigate($N=1$ case), whether a Hopf term
,associated with this new soliton number, can be added to the model to
obtain fractional spin. The question is all the more important, as the model
has already got an nonabelian $SU(2)$ CS term, needed for the {\it very existence}
of these new type of solitons. And, as we have mentioned earlier, the CS
term, is likely to play its own role in imparting fractional spin to the
model. We find that the fractional angular momentum is given in terms of
both the soliton number and the nonabelian charge.

To that end, we organise the paper as follows. In section 2, we carry out
the Hamiltonian analysis of the gauged $CP^1$ model. The Hopf term is introduced
in section 3 and again the Hamiltonian analysis of the resulting model is
performed. In section 4, we compute the fractional spin of the model.
Finally we conclude in section 5.

2.{\bf Hamiltonian Analysis of Gauged $CP^1$ model}

We are going to carry out the Hamiltonian analysis of the gauged
$CP^1$ model, as introduced by Cho and Kimm[19]. The model is
given by,
$${\cal L}=(D_\mu Z)^{\dg}(D^{\mu}Z)+\t \e^{\mu \nu \l}
[{A^a}_{\mu}\p_{\nu}{A^a}_{\l}+{g\over 3}\e^{abc}A^a_{\mu}A^b_{\nu}A^c_{\l}]
-{\l}(Z^{\dg}Z-1)\eqno(2.1)$$
where $Z=\pmatrix{z_1 \cr z_2}$ is an $SU(2)$ doublet satisfying
$$Z^{\dg}Z=1\eqno(2.2)$$
and enforced by the Lagrange multiplier $\l$ in (2.1). The
covariant derivative operator $D_{\mu}$ is given as
$$D_{\mu}=\p_{\mu}-ia_{\mu}-igA^a_{\mu}T^a \eqno(2.3)$$
with $T^a={1\over 2}{\s}^a$ (${\s}^a$, the Pauli matrices), representing
the $SU(2)$ generators and $g$ a coupling constant.$\t$ represents the
CS parameter.And finally $a_{\mu}$ and $A_{\mu}^a$ represent the
$U(1)$ and $SU(2)$ gauge fields respectively. Note that there is no
dynamical CS term for the the $a_{\mu}$ field.

The canonically conjugate momenta variables corresponding to the
configuration space variables ($a_{\mu}$,$A_{\mu}^a$,$z_{\a}$,$z_{\a}^*$)
are given as,
$$\pi^{\mu}={\d L\over {\d \dot {a_{\mu}}}}=0$$
$$\Pi^{ia}={\d L\over {\d \dot {A^a_i}}}=\t \e^{ij}A^a_j;
\Pi^{0a}={\d L\over {\d \dot {A^a_0}}}=0\eqno(2.4)$$
$$\pi_{\a}={\d L\over {\d \dot z_{\a}}}=(D_0Z)^*_{\a};
\pi^*_{\a}={\d L\over {\d \dot z_{\a}^*}}=(D_0Z)_{\a}$$
where $L=\int d^2x {\cal L}$ is the Lagrangian.

The Legendre transformed Hamiltonian 
$${\cal H}=\pi^{\mu}{\dot a_{\mu}}+\Pi^{\mu a}{\dot A^a_{\mu}}+\pi_{\a}
{\dot z}_{\a}+\pi_{\a}^*{\dot z}^*_{\a}-\cal L \eqno(2.5a)$$
when expreseed in terms of the phase space variables (2.4), gives
$${\cal H}=\pi_{\a}^*\pi_{\a} +ia_0(\pi_{\a}z_{\a}-\pi_{\a}^*z_{\a}^*)+
{1\over 2}igA^a_0[\pi_{\a}(\s^aZ)_{\a}-\pi^*_{\a}(Z^{\dg}\s^a)_{\a}]$$
$$-2 \t A^a_0B^a+(D_iZ)^{\dg}(D_iZ)+\l (Z^{\dg}Z-1)\eqno(2.5b)$$
where
$$B^a\equiv F^a_{12}=(\p_1A_2^a-\p_2A^a_1+g\e^{abc}A_1^bA_2^c)\eqno(2.5c)$$
is the non-abelian $SU(2)$ magnetic field.

Clearly the fields $a_0$,$A^a_0$ and $\l$ play the role of Lagrange
multipliers, which enforce the following constraints,
$$G_1(x) \equiv i(\pi_{\a}(x)z_{\a}(x)-\pi_{\a}^*(x)z_{\a}^*(x))\approx 0\eqno(2.6)$$
$$G_2^a(x) \equiv {ig \over 2}[\pi_{\a}(x)(\s^aZ(x))_{\a}-\pi^*_{\a}(x)(Z^{\dg}(x)\s^a)_{\a}]-2 \t B^a(x) \approx 0\eqno(2.7)$$
$$\chi_1(x) \equiv Z^{\dg}(x)Z(x)-1 \approx 0.\eqno(2.8)$$
Apart from all these constraints, we have yet another primary constraint,
$$\chi_2(x) \equiv (\pi_{\a}(x)z_{\a}(x)+\pi_{\a}^*(x)z_{\a}^*(x))\approx 0\eqno(2.9)$$
Also the preservation of the primary constraint $\pi^i(x) \equiv 0$ (2.4)
yield the following secondary constraint,
$$2iZ^{\dg}\p_jZ + 2a_j + gA_j^aM^a \approx 0 \eqno(2.10)$$
Here
$$M^a=Z^{\dg}\s^aZ\eqno(2.11)$$ 
is a unit 3-vector, obtained from the $CP^1$ variables using the Hopf map.
We are left with a pair of primary constraints from the CS gauge field
sector in (2.4),
$$\xi^{ia}\equiv \Pi^{ia}-\t \e^{ij}A^a_j\approx 0\eqno(2.12)$$
This pair of constraints can be implemented strongly by the bracket,
$$\{A^a_i(x),A^b_j(y)\}={1\over 2 \t}\e_{ij}\d^{ab}\d (x-y)\eqno(2.13)$$
obtained either by using 
Dirac method[20] or by the symplectic technique of Faddeev-Jackiw[21].

Also note that the constraint $\pi^i\approx 0$ (2.4) is conjugate to the
constraint (2.10) and can again be strongly implemented by the Dirac bracket
(DB),
$$\{\pi^i(x),a_j(y)\}=0\eqno(2.14)$$
With this the `weak' equality in (2.10) is actually rendered into a strong
equality and the field $a_i$ ceases to be an independent degree of freedom.

Finally the constraints $\chi_1$ (2.8) and $\chi_2$ (2.9) are conjugate
to each other and are implemented strongly by the following DBs,
$$\{z_{\a}(x),z_{\b}(y)\}= \{z_{\a}(x),z_{\b}^*(y)\}=0$$
$$\{z_{\a}(x),\pi_{\b}(y)\}=(\d_{\a \b}-{1\over 2}z_{\a}z_{\b}^*)\d (x-y)$$
$$\{z_{\a}(x),\pi_{\b}^*(y)\}=-{1\over 2}z_{\a}z_{\b}\d (x-y)\eqno(2.15)$$
$$\{\pi_{\a}(x),\pi_{\b}(y)\}=-{1\over 2}(z_{\a}^*\pi_{\b}-z_{\b}^*\pi_{\a})\d (x-y)$$
$$\{\pi_{\a}(x),\pi_{\b}^*(y)\}=-{1\over 2}(z_{\a}^*\pi_{\b}^*-z_{\b}\pi_{\a})\d (x-y)$$
Precisely the same set of brackets (2.15) are obtained in the case of $CP^1$
model also[22]. We are thus left with the constraints (2.6) and (2.7) and are
expected to be the Gauss constraints generating $U(1)$ and $SU(2)$ gauge
transformations respectively. The fact that this is indeed true will be
exhibited by explicit computations. But before we proceed further, let
us note that the constraints (2.8),(2.9) and (2.10) hold strongly now. In
view of this, the constraint $G_1$ (2.6) can be simplified as,
$$G_1(x)=2i\pi_{\a}(x)z_{\a}(x) \approx 0 \eqno(2.16)$$
At this stage, one can substitute $\pi_{\a}=(D_0Z)^{\dg}_{\a}$ from (2.4)
and solve for $a_0$ to get,
$$a_0=-iZ^{\dg}\p_0Z-{1\over 2}gA^a_0M^a \eqno(2.17)$$
Clearly this is not a constraint equation, as it involves time derivative.
It is nevertheless convenient to club it with the expression of $a_i$,
obtained from (2.10) and write covariantly as,
$$a_{\mu}=-iZ^{\dg}\p_{\mu}Z-{1\over 2}gA^a_{\mu}M^a \eqno(2.18)$$
Here the first term $(-iZ^{\dg}\p_{\mu}Z)$ is the pullback, onto the
spacetime, of the $U(1)$ connection on the $CP^1$ manifold[16]. The second 
term on the other hand has nothing to do with $CP^1$ connection and arises
from the presence of the CS gauge field $A^a_{\mu}$.

It is now quite trivial to show that $G_1(x)$ (2.16) generates $U(1)$ gauge
transformation on the $Z$ fields
$$\d Z(x)=\int d^2y f(y)\{Z(x),G_1(y)\}=if(x)Z(x)\eqno(2.19)$$
but leaves the CS gauge field $A^a_{\mu}$ unaffected
$$\d A^a_{\mu}(x)=\int d^2y f(y)\{A^a_{\mu}(x),G_1(y)\}=0\eqno(2.20)$$
Consequently $M^a=Z^{\dg}\s^aZ$ (2.11) remains invariant under this
transformation and hence $a_{\mu}$ (2.18) undergoes the usual gauge
transformation
$$\d a_{\mu}(x)=\int d^2y f(y)\{a_{\mu}(x),G_1(y)\}=\p_{\mu}f(x) \eqno(2.21)$$
Here in the equations (2.19-2.21) we have taken $f(x)$ to be an arbitrary
differentiable functions with compact support.

Proceeding similarly, one can show that the constraints $G_2^a(x)$ (2.7)
generates $SU(2)$ gauge transformation,
$$\d Z(x)\equiv \int d^2y f^a(y)\{Z(x),G^a_2(y)\}=ig f^a(x)(T^aZ(x))\eqno(2.22a)$$
$$\d A^a_i(x)\equiv \int d^2y f^b(y)\{A^a_i(x),G^b_2(y)\}=
\p_if^a(x)-g\e^{abc}f^b(x)A^c_i(x)\eqno(2.22b)$$
Using these one can also show that,
$$\d F^a_{ij}=-g\e^{abc}f^bF^c_{ij}$$
$$\d M^a=-g\e^{abc}f^bM^c\eqno(2.23a)$$
but $M^aF^a_{ij}$ is an $SU(2)$ scalar as
$$\d (M^aF^a_{ij})=0\eqno(2.23b)$$
It also follows from (2.23a) and (2.18) that $a_{\mu}$ remains unaffected
by this $G_2^a$,
$$\d a_{\mu}(x)=\int d^2y f^a(y)\{a_{\mu}(x),G^a_2(y)\}=0\eqno(2.24)$$
just as $A_{\mu}^a(x)$ remains unaffected by $G_1$ (2.20).

The fact that $G_1(x)$ and $G_2^a(x)$ are indeed  the first class constraints
of the model can be easily seen. Firstly one has to just rewrite $G_1$(2.16)
using (2.4) as
$$G_1(x)=2i(D_0Z)^{\dg}Z \approx 0\eqno(2.25)$$
to see that this is manifestly invariant under the $SU(2)$ gauge transformation
generated by $G_2^a(x)$ (2.22). We thus have,
$$\{G_1(x),G^a_2(y)\}=0\eqno(2.26)$$
It also follows after a straightforward algebra that $G^a_2$'s satisfy an
algebra isomorphic to $SU(2)$ Lie algebra and thus vanishes on the constraint
surface,
$$\{G^a_2(x),G^b_2(y)\}=2\e^{abc}G_2^c(x)\d (x-y)\approx 0\eqno(2.27)$$

Finally note that (2.18) really corresponds to the Euler-Lagrange's equation
for the $a_{\mu}$ field. The corresponding equations for $Z$ and $A^a_{\l}$
are given by,
$$D_{\mu}D^{\mu}Z + \l Z =0 \eqno(2.28)$$
$$\t \e^{\mu \nu \l}F_{\nu \l}^a=ig[(D^{\mu}Z)^{\dg}T^aZ- Z^{\dg}T^a(D^{\mu}Z)]\eqno(2.29)$$
respectively.

3.{\bf Introducing the Hopf term}

In order to introduce the Hopf term, it will be convenient to provide
a very brief review of some of the essential features of these new solitons.
For this we essentially follow [19].
The symmetric expression for the energy-momentum(EM) tensor, as obtained by
functionally differentiating the action $S(=\int d^3x {\cal L})$ with respect
to the metric, is given by
$$T_{\mu \nu}=(D_{\mu}Z)^{\dg}(D_{\nu}Z)+(D_{\nu}Z)^{\dg}(D_{\mu}Z)
-g_{\mu \nu}(D_{\rho}Z)^{\dg}(D^{\rho}Z)\eqno(3.1)$$
The energy functional 
$$E=\int d^2x T_{00}=\int d^2x [2(D_0Z)^{\dg}(D_0Z)-(D_{\mu}Z)^{\dg}(D^{\mu}Z)]\eqno(3.2)$$
can be expressed alternatively as,
$$E=\int d^2x ( |D_0Z|^2+ |(D_1\pm iD_2)Z|^2) \pm 2\pi N \eqno(3.3a)$$
where 
$$N={1\over 2\pi i}\int d^2x \e^{ij}(D_iZ)^{\dg}(D_jZ)\eqno(3.3b)$$
is the soliton charge.

It immediately follows that the energy functional satisfy the following
inequality,
$$E \geq 2\pi |N| \eqno(3.4)$$
The corresponding saturation conditions are,
$$|D_0Z|^2=|(D_1\pm iD_2)Z|^2=0 \eqno(3.5)$$
For static configuration ($\dot Z=0$), this yields
$$A^a_0=k M^a\eqno(3.6)$$
where $k$ is an arbitrary constant.

Again assuming the static case, one can easily show that $\mu =0$ component
of the Euler-Lagrange equation (2.29) implies that the $SU(2)$ magnetic
field $B^a$ vanishes,
$$B^a=0 \eqno(3.7)$$
where use of (3.6) has been made. This in turn implies that $A^a_i$ is a
pure gauge, so that one can write without loss of generality
$$A^a_i=0 \eqno(3.8)$$
In this gauge, the soliton charge $N$ (3.3b) reduces to the standard $CP^1$
soliton charge,
$$N={1\over 2\pi i}\int d^2x \e^{ij}({\cal D}_iZ)^{\dg}({\cal D}_jZ)\eqno(3.9a)$$
where
$${\cal D}_i=D_i|_{A_i^a=0}=\p_i-(Z^{\dg}\p_iZ)\eqno(3.9b)$$
is the covariant derivative operator for the standard $CP^1$ model. Thus
in this gauge (3.8), the ``soliton charge" is essentially characterised by
$\pi_2(CP^1)=Z$. Nonetheless, it is possible to make ``large'' topology
changing gauge transformation, where $A^a_i$ is no longer zero and one
{\it has} to make use of (3.3b), rather than (3.9a), to compute the
solitonic charge. Of course this will yield the same value for the charge,
but the various solitonic sectors will not be characterised by $\pi_2(CP^1)$
anymore.

To make things explicit, consider a typical solitonic configuration:
$$Z={1\over {\sqrt {r^2+\l^2}}}\pmatrix{re^{-i\Phi} \cr \l}\eqno(3.10a)$$
$$A_i^a=0\eqno(3.10b)$$
where ($r$,$\Phi$) represents the polar coordinates in the two-dimensional
plane and $\l$ is the size of the soliton. The corresponding unit vector
$M^a$(2.11) takes the form,
$$M^1=sin \T cos \Phi ={2r\l \over {r^2+ \l^2}}cos \Phi$$
$$M^2=sin \T sin \Phi ={2r\l \over {r^2+ \l^2}}sin \Phi \eqno(3.11)$$
$$M^3=cos \T ={r^2-\l^2 \over {r^2+\l^2}}$$
We therefore have for the time component of the gauge field $A_0^a=kM^a$(3.6).

At this stage, one can make a ``large'' topology changing gauge transformation,
$$Z \rightarrow Z'=UZ=\pmatrix{0 \cr 1}\eqno(3.12a)$$
where,
$$U={1\over {\sqrt {r^2+\l^2}}}\pmatrix{\l & -re^{-i\Phi} \cr re^{i\Phi} & \l}\in SU(2)\eqno(3.12b)$$
so that $A^a_0$ undergoes the transformation,
$$A^a_0\rightarrow {A'}^a_0=-k\pmatrix{0 \cr 0 \cr 1}\eqno(3.12c)$$
The spatial components on the other hand, undergoes the transformation,
$$A_i\rightarrow A'_i=UA_iU^{-1}+{i\over g}U\p_iU^{-1}=-{i\over g}(\p_iU)U^{-1}$$
which on further simplification yields the following form for the connection
one-form in the cartesian coordinate system,
$${A'}^1={2\l \over g(r^2+\l^2)}dy$$
$${A'}^2=-{2\l \over g(r^2+\l^2)}dx\eqno(3.12d)$$
$${A'}^3=-{2 \over g(r^2+\l^2)}(xdy-ydx)$$
Now it is a matter of straightforward exercise to calculate the soliton charge
`$N$' in either of these gauges (3.10) and (3.12). For example, in the gauge
(3.10), this can be computed by using (3.9a) to get,
$$N={1\over {2\pi}}\int d(-iZ^{\dg}dZ)=-1 \eqno(3.13)$$
On the other hand, the same soliton charge can also be computed in the gauge
(3.12), but where the use of (3.3b), rather than (3.9a), has to be made. Note
that the topological density $j^0$ ($N\equiv \int d^2x j^0$) can be written as,
$$j^0={1\over 2\pi i}\e^{ij}(D_iZ)^{\dg}(D_jZ)={\tilde j}^0+{g\over 4\pi}
\e^{ij}A^a_i(\p_jM^a+{g\over 2}\e^{abc}A^b_jM^c)\eqno(3.14a)$$
where,
$${\tilde {j}}^0={\e^{ij}\over 2\pi i}({\cal D}_iZ)^{\dg}({\cal D}_jZ)\eqno(3.14b)$$
is the expression of topological density in the gauge (3.10). But in the gauge (3.12),
this ${\tilde j}^0$ vanishes, and one can rewrite $N$ completely in terms of the
CS gauge field as,
$$N={g^2\over {8\pi k}}\int d^2x \e^{ij} \e^{abc}A^a_0A^b_iA^c_j \eqno(3.15)$$
The corresponding $Z$ field configuration being trivial ($Z=\pmatrix{0 \cr 1}$)
, the soliton number $N$ cannot be captured by $\pi_2(CP^1)$. 

Since `$N$' is a conserved soliton charge, with an associated topological
density $j^0$(3.14a), one can regard $j^0$ to be the time-component of a
conserved topological $3(=2+1)$-current 
$$j^{\mu}={1\over {2\pi i}}\e^{\mu \nu \l}(D_{\nu}Z)^{\dg}(D_{\l}Z)\eqno(3.16)$$ 
This can therefore be expressed as the curl of a `fictitious' $U(1)$ 
gauge field ${\cal A}_{\l}$:
$$j^{\mu}={1\over {2\pi}}\e^{\mu \nu \l}\p_{\nu}{\cal A}_{\l}\eqno(3.17)$$
Unlike the case of pure $CP^1$ model[11,12], this equation cannot be solved
trivially for ${\cal A}_{\l}$ in a gauge independent manner[2,16]. We therefore
find it convenient to follow Bowick et.al.[13], to solve (3.17) for ${\cal A}_{\l}$
in the radiation gauge ($\p_i{\cal A}_i=0$), where one can prove the
following identity,
$$\int d^3x j_0{\cal A}_0=- \int d^3x j_i{\cal A}_i\eqno(3.18)$$
so that the Hopf action
$$S_{Hopf}=\T \int d^3x j^{\mu}{\cal A}_{\mu}, \eqno(3.19)$$
($\T $ being the Hopf parameter and should not be confused with the spherical
angles introduced in (3.11)) simplifies to the following non-local term
$$S_{Hopf}=-2 \T \int d^3x j_i{\cal A}_i. \eqno(3.20)$$
Adding this term to the original model (2.1), we get the following model,
$${\cal L}=(D_\mu Z)^{\dg}(D^{\mu}Z)+\t \e^{\mu \nu \l}
[{A^a}_{\mu}\p_{\nu}{A^a}_{\l}+{g\over 3}\e^{abc}A^a_{\mu}A^b_{\nu}A^c_{\l}]$$
$$+{\T \over \pi i}\e^{ij}{\cal A}_i[(D_jZ)^{\dg}(D_0Z)-(D_0Z)^{\dg}(D_jZ)]
 -{\l}(Z^{\dg}Z-1)\eqno(3.21)$$
In the rest of this section, we shall be primarily concerned with
the Hamiltonian analysis of this model. As the Hopf term is linear in time
derivative of the $Z$ variable, the analysis is expected to undergo only minor 
modification. Indeed we shall verify this by explicit computations.

To begin with, note that the only change in the form of canonically conjugate
momenta variables takes place in the variables $\tilde {\pi}_{\a}$ and its
complex conjugates, counterpart of  $\pi_{\a}$ and $\pi_{\a}^*$ (2.4)-the
momenta variables for the model (2.1). They are now given as,
$$\tilde {\pi}_{\a}=(D_0Z)^*_{\a}+{\T \over {\pi i}}\e^{ij}{\cal A}_i(D_jZ)^*_{\a}
=\pi_{\a}+{\T \over {\pi i}}\e^{ij}{\cal A}_i(D_jZ)^*_{\a}$$
$$\tilde {\pi}^*_{\a}=(D_0Z)_{\a}-{\T \over {\pi i}}\e^{ij}{\cal A}_i(D_jZ)_{\a}
=\pi^*_{\a}-{\T \over {\pi i}}\e^{ij}{\cal A}_i(D_jZ)_{\a}\eqno(3.22)$$
 Rest of the momenta variables undergo no change from that
of (2.4).

The Legendre transformed Hamiltonian $\tilde {\cal H}$ can be calculated in
a straightforward manner to get,
$$\tilde {\cal H}={\cal H}+{g\T \over 2\pi}A^a_0\e^{ij}{\cal A}_i(D_jM)^a\eqno(3.23)$$
where $\cal H$ is just the expression of the Legendre transformed Hamiltonian
density (2.5) corresponding to the model (2.1) and $(D_jM)^a$ is given by,
$$D_jM^a=\p_jM^a+g\e^{abc}A^b_jM^c\eqno(3.24)$$
as can be easily obtained by using the Hopf map (2.11) and the fact that the
covariant derivative operator $D_{\mu}$ boils down, using (2.3) and (2.18) to,
$$D_{\mu}Z=\p_{\mu}Z-(Z^{\dg}\p_{\mu}Z)Z+{ig \over 2}A^a_{\mu}(M^a-\s^a)Z\eqno(3.25)$$
Clearly the structure of all the constraints remain the same, except the $SU(2)$
Gauss constraint. This is clearly given as,
$${\tilde G}^a_2=G^a_2+{g\T \over 2\pi}\e^{ij}{\cal A}_i(D_jM)^a \eqno(3.26)$$
where $G^a_2$ is given in (2.7). But we have to rewrite this in terms of 
${\tilde \pi}_{\a}$ and ${\tilde \pi}^*_{\a}$. Once we do this, we find that
that the $SU(2)$ Gauss constraint $G^a_2$ (2.7) for the model (2.1) is now
given by,
$$G^a_2=ig\Bigl([{\tilde {\pi}}_{\a}(T^aZ)_{\a}-{\tilde {\pi}}^*_{\a}(Z^{\dg}T^a)_{\a}]
+{i\T \over {2\pi}}\e^{ij}{\cal A}_i(D_jM)^a\Bigr)-2\t B^a\approx 0$$
Substituting this in (3.26), ${\tilde G}^a_2$ is found to have the same form
as that of $G^a_2$ (2.7) with the replacement $\pi_{\a}\rightarrow {\tilde \pi}_{\a}$
and $\pi^*_{\a}\rightarrow {\tilde \pi}^*_{\a}$,
$${\tilde G}^a_2=ig({\tilde {\pi}}_{\a}(T^aZ)_{\a}
-{\tilde {\pi}}^*_{\a}(Z^{\dg}T^a)_{\a})-2\t B^a\approx 0\eqno(3.27)$$
The other $U(1)$ Gauss constraint ${\tilde G}_1$ (2.6) can also be seen to take
the same form, with identical replacement,
$${\tilde G}_1(x)=i({\tilde \pi}_{\a}(x)z_{\a}(x)
-{\tilde \pi}^*_{\a}{z}^*_{\a}(x))\approx 0\eqno(3.28)$$
, where use of the identity $Z^{\dg}D_{\mu}Z=0$ has been made.

Finally note that the constraint (2.9) also preserves its form,i.e.
$$\chi_2(x)={\tilde \pi}_{\a}(x)z_{\a}(x)+ c.c $$
and is again conjugate to $\chi_1$ (2.8). Thus these pair of constraints
can be implemented strongly by using the DB (2.15), again taken with the
replacement $\pi_{\a}\rightarrow {\tilde \pi}_{\a}$ and
$\pi^*_{\a}\rightarrow {\tilde \pi}^*_{\a}$. On the other hand, the pair
of second class constraints (2.12) are implemented strongly by the brackets
(2.13) in this case also. These set of DB furnishes us with the symplectic
structure of the model (3.21).

4. {\bf Angular momentum}

In this section, we are going to find the fractional spin imparted by
the Hopf term. As was done for the models involving the CS[7-9] and Hopf[13]
term, the fractional spin was essentially revealed by computing the
difference $(J^s-J^N)$ between the expression of angular momentum $J^s$,
obtained from the symmetric expression of the EM tensor $T_{\mu \nu}$
($\sim {\d S\over \d g^{\mu \nu}}$) and the one $J^N$, obtained by using
Noether's prescription. It is $J^s$, which is taken to be the physical
angular momentum. This is because it is gauge invariant by construction,
in contrast to $J^N$, which turn out to be gauge invariant only on the
Gauss constraint surface and that too usually under those gauge transformations,
which tend to identity asymptotically[8,9].

To that end, let us consider the generator of linear momentum. This is
obtained by integrating the (0i) component of the EM tensor (3.1), which
undergoes no modification as the metric independent topological (Hopf)
term (3.20) is added to the original Lagrangian (2.1) to get the model (3.21).
$$P_i^s=\int d^2x T^s_{0i}=\int d^2x[(D_0Z)^{\dg}(D_iZ)+(D_iZ)^{\dg}(D_0Z)]\eqno(4.1)$$
Expressing this in terms of phase-space variables (3.22), one gets
$$P_i^s=\int d^2x[{\tilde \pi}_{\a}(D_iZ)_{\a}+{\tilde \pi}^*_{\a}(D_iZ)^*_{\a}
+2 \T{\cal A}_i(x)j^0(x)]\eqno(4.2)$$
This can now be re-expressed as,
$$P_i^s=\int d^2x [{\tilde \pi}_{\a}\p_iz_{\a}+{\tilde \pi}^*_{\a}\p_iz^*_{\a}
-2\t A^a_iB^a+2\T{\cal A}_ij^0-a_i{\tilde G}_1-A^a_i{\tilde G}^a_2]\eqno(4.3)$$
However this cannot be identified as an expression of linear momentum, as this
fails to generate appropriate translation,
$$\{Z(x),P^s_i\}\approx D_iZ \eqno(4.4)$$
in contrast to the corresponding expression of linear momentum 
$$P_k^N=\int d^2x T^N_{0k}=\int d^2x [{\tilde \pi}_{\a}\p_kz_{\a}
+{\tilde \pi}^*_{\a}\p_kz^*_{\a}-\t \e^{ij}A^a_i\p_kA^a_j]\eqno(4.5)$$
obtained through Noether's prescription, as this generates appropriate
translation by construction,
$$\{Z(x),P^N_k\}=\p_kZ(x)$$
$$\{A^a_i(x),P^N_k\}=\p_kA^a_i\eqno(4.6)$$
The adjective ``appropriate'' in this context means that the bracket
$\{\Phi (x),{\cal G}\}$ is just equal to the Lie derivative
(${\cal L}_{V_{\cal G}}({\Phi (x)})$) of a generic field $\Phi (x)$ 
with respect to the vector field $V_{\cal G}$, associated to the symmetry 
generator ${\cal G}$. We have not,of course, displayed any indices here.
The field $\Phi$ may be a scalar, spinor, vector or tensor field in
general. In this case, it
can correspond either to the scalar field $Z(x)$ or the vector field
$A^a_i(x)$.
And ${\cal G}$ can be, for example, the momentum($P_i$) or angular momentum
($J$)operator generating translation and spatial rotation respectively. The
associated vector fields $V_{\cal G}$ are thus  given as $\p_i$ and $\p_{\phi}$
respectively ($\phi$ being the angle variable in the polar coordinate system
in 2-dimensional plane).

Coming back to the translational generator $P_i^s$ (4.3), we observe that
the EM tensor (3.1) is not unique by itself. One has the freedom to modify
it to ${\tilde T}_{\mu \nu}$ by a linear combination of first class constraint(s), here
${\tilde G}_1$ (3.28) and ${\tilde G}^a_2$ (3.27) with arbitrary tensor valued
coefficients $u_{\mu \nu}$ and $v^a_{\mu \nu}$:
$${\tilde T}_{\mu \nu}=T_{\mu \nu}+u_{\mu \nu}{\tilde G}_1
+v^a_{\mu \nu}{\tilde G}^a_2\eqno(4.7)$$
Choosing,
$$u_{0i}=a_i$$
$$v^a_{0i}=A^a_i \eqno(4.8)$$
one can easily see that the corresponding modified expression of momentum
$${\tilde P}_i=\int d^2x {\tilde T}_{0i}=\int d^2x [{\tilde \pi}_{\a}\p_iz_{\a}
+{\tilde \pi}^*_{\a}\p_iz^*_{\a}-2\t A^a_iB^a+2\T {\cal A}_ij_0]\eqno(4.9)$$
generate appropriate translation,
$$\{Z(x),{\tilde P}_i\}=\p_iZ(x)$$
$$\{A^a_k(x),{\tilde P}_i\}=\p_iA^a_k(x)\eqno(4.10)$$
just like $P_i^N$ (4.4).

So finally the corresponding expression of angular momentum can be written
as,
$$J^s=\int d^2x \e^{ij}x_i{\tilde T}_{0j}
=\int d^2x \e^{ij}x_i[{\tilde \pi}_{\a}\p_jz_{\a}
+{\tilde \pi}^*_{\a}\p_jz^*_{\a}-2\t A^a_jB^a+2\T {\cal A}_jj_0]\eqno(4.11)$$
$$J^N=\int d^2x [\e^{ij}x_i({\tilde \pi}_{\a}\p_jz_{\a}
+{\tilde \pi}^*_{\a}\p_jz^*_{\a}-\t \e^{kl}A^a_k\p_jA^a_l)-\t A^a_jA^a_j]
\eqno(4.12)$$
Just like the case the case of linear momentum, here too one can show that
both $J^s$ and $J^N$ generate appropriate spatial rotation,
$$\{Z(x),J^s\}=\{Z(x),J^N\}=\e^{ij}x_i\p_jZ(x)$$
$$\{A_k^a(x),J^s\}=\{A^a_k(x),J^N\}=\e^{ij}x_i\p_jA^a_k(x)+\e_{ki}A^{ai}(x)\eqno(4.13)$$
(Again the adjective ``appropriate" has been used in the sense, mentioned
above.)
However, they are not identical and the difference $J_f\equiv (J^s-J^N)$ is
given as,
$$J_f=\t \int d^2x \p_i[x_jA^{aj}A^{ai}-x^iA^a_jA^{aj}]
+2\T \int d^2x \e^{ij}x_i{\cal A}_jj_0 \eqno(4.14)$$
The first $\t$-dependent boundary term has occured earlier in [8,9], where
some of its properties were studied in detail. For example, it was noted that
this term is gauge invariant under only those gauge transformation which
tends to identity asymptotically[9]. To evaluate it in a rotationally symmetric
configuration therefore, one can make use of the radiation gauge
$(\p_iA_i^a=0)$ condition. To this end, let us rewrite the Gauss constraint
(3.27) as,
$$j^a_0\approx 2\t B^a \eqno(4.15a)$$
with $j^a_0=ig({\tilde \pi}_{\a}(T^aZ)_{\a}-{\tilde \pi}^*_{\a}(Z^{\dg}T^a)_{\a})$
being the time component of the conserved $3(=2+1)$-current
$$j^{a\mu}=ig[(D^{\mu}Z)^{\dg}T^aZ-Z^{\dg}T^a(D^{\mu}Z)]$$
$$+{\T g\over \pi}\e^{ij}{\cal A}_i[(Z^{\dg}T^a(D_jZ)+(D_jZ)^{\dg}T^aZ)\d^{\mu}_0-
(Z^{\dg}T^a(D_0Z)+(D_0Z)^{\dg}T^aZ)\d^{\mu}_j]\eqno(4.15b)$$
arising from the global $SU(2)$ symmetry of the model (3.21). Note that
we have expressed this in terms of the phase space variables.

On integrating (4.15a), one gets 
$$Q^a\approx 2\t \Phi^a \eqno(4.16)$$
where $Q^a(\equiv \int d^2x j^a_0)$ and $\Phi^a(\equiv \int d^2x B^a)$
represent the $SU(2)$ charge and magnetic flux respectively. Proceeding as
in [8], we can write the following configuration of the $SU(2)$ gauge field,
$$A^a_i=-{Q^a\over 2\pi \t}\e_{ij}{x^j\over r^2}\eqno(4.16)$$
in the radiation gauge ($\p_iA^a_i=0$). Using this, one can easily show that
the first $\t$-dependent term in (4.14) yields ${Q^aQ^a\over 2\pi \t}$ and
the second $\T$-dependent term yields, following [13], $\T N^2$. So finally,
we have from (4.14),
$$J_f={Q^aQ^a\over 2\pi \t}+\T N^2 \eqno(4.17)$$
We thus see that the classical expression of fractional angular momentum (4.17)
contains two terms. One depends on the soliton number $N$ and the other on
the nonabelian charge $Q^a$. The former is just
as in the model, where Hopf term is coupled to NLSM [3,13]. On the other hand,
the latter is a typical nonabelian expression, as in [8].

5. {\bf Conclusions}

In this paper, we have carried out the classical Hamiltonian analysis of
the gauged $CP^1$ model of Cho and Kimm[19]. As was shown in [19], the
model has got its own solitons, the very existence of which depends
crucially on the presence of $SU(2)$ CS term. These solitons are
somewhat more general then that
of NLSM[10]. We use the adjective ``general'' to indicate that these
solitons can be characterised by ${\pi_2(CP^1)}=Z$ only for the gauge
$A_i^a=0$ (Note that $A_i^a$ is a pure gauge). One can make large topology
changing gauge transformation  and thereby making $A^a_i\neq 0$, without
changing the soliton number. We then constructed the Hopf term associated
to these solitons and again carried out the Hamiltonian analysis of the
model(3.21), obtained by adding Hopf term to (2.1), to find that
the symplectic structure and the structure of the 
constraints undergo essentially no modification, despite the fact that
the form of the momenta variables conjugate to $z_{\a}$ and $z^*_{\a}$
undergo changes. We then calculated the fractional angular momentum
by computing the difference between $J^s$ and $J^N$, the expressions
of angular momenta obtained from the symmetric expression of energy-momentum
tensor and the one obtained through Noether's prescription respectively.
We find that this fractional angular momentum consists of two pieces,
one is given in terms of the soliton number and the other is given in terms
of the nonabelian ($SU(2)$) charge. In absence of the
Hopf term $(\T =0)$ (i.e. for the model (2.1)), only this latter term will
contribute. Again as in [8], this term can be shown to consist of two
pieces, one which involves a direct product in the isospin space and
characterises a typical nonabelian feature, while the other contains
the abelian charge defined in a nonabelian theory.

\pagebreak

{\bf References}

\begin{enumerate}

\item F.Wilczek (Ed.)``Fractional Statistics and Anyonic Superconductivity'',
(World Scientific, Singapore,1990)

\item S.Forte, Rev.Mod.Phys.{\bf 64}(1992)193.

\item F.Wilczek and A.Zee, Phys.Rev.Lett.{\bf 51}(1983)2250.

\item A.P.Balachandran, G.Marmo, B.S.Skagerstam and A.Stern,``Classical
Topology and Quantum States'',(World Scientific, Singapore,1991).

\item C.R.Hagen, Ann.Phys.(N.Y.){\bf 157}(1984)342.

\item P.Panigrahi, S.Roy and W.Scherer, Phys.Rev.Lett. {\bf 61}(1988)2827.

\item R.Banerjee, Phys. Rev. Lett. {\bf 69} (1992) 17; Phys. Rev. {\bf
D41} (1993) 2905; Nucl. Phys. {\bf B390} (1993) 681; R.Banerjee and
B.Chakraborty, Phys. Rev. {\bf D49} (1994) 5431.

\item R.Banerjee and B.Chakraborty, Ann.Phys.(N.Y.){\bf 247}(1996)188.

\item B.Chakraborty, Ann. Phys. (N.Y.) {\bf 244} (1995) 312; B.
Chakraborty and A. S. Majumdar, Ann. Phys. {\bf 250} (1996) 112.

\item R.Rajaraman,``Solitons and Instantons''(North Holland,1982);C.Rebbi and
G.Soliani,``Solitons and Particles''(World Scientific, Singapore,1984);
V.A.Novikov, M.A.Shifman, A.I.Vainshtein and V.I.Zhakharov, Phys.Rep.{\bf 116}
(1984)103.

\item Y.S.Wu and A.Zee, Phys.Lett. {\bf 147B}(1984)325.

\item B.Chakraborty and T.R.Govindarajan, Mod.Phys.Lett.{\bf A12}(1997)619.

\item M.Bowick,D.Karabali and L.C.R.Wijewardhana, Nucl.Phys.{\bf B271}(1986)417.

\item J.Zinn-Justin,``Quantum Field Theory and Critical Phenomena''
(Clarendon Press,Oxford, 1990); E.Fradkin,``Field Theory of Condensed matter
systems''(Addison-Wesley,1991).

\item L.Landau and E.Lifshitz, Phys. A. {\bf 8} (1935) 153; A. Kosevich,
B. Ivanov and A. Kovalev, Phys. Rep. {\bf 194} (1990) 117.

\item R.Banerjee and B.Chakraborty, Nucl.Phys.{\bf B449}(1995)317.

\item H.Kobayashi, I.Tsutsui and S.Tanimura,``Quantum Mechanically Induced
Hopf Term in the $O(3)$ non-linear sigma model'', hep-th/9705183.

\item G.Nardelli, Phys.Rev.Lett. {\bf 73}(1994)2524.

\item Y.Cho and Kimm, Phys.Rev.{\bf D52}(1995)7325.

\item P.A.M.Dirac,``Lectures on Quantum Mechanics'', Belfar Graduate School of
Science, Yeshiva University, New York, 1964; A.Hanson, T.Regge and C.Teitelboim,
``Constraint Hamiltonian Analysis'', (Academia Nazionale dei Lincei, Rome,1976).

\item L.Faddeev and R.Jackiw, Phys.Rev.Lett.{\bf 60}(1988)1692.

\item R.Banerjee, Phys.Rev.{\bf D49}(1994)2133.

\end{enumerate}

\end{document}